\documentclass[aps,prd,twocolumn,10pt,tightenlines,superscriptaddress,
nofootinbib,showpacs]{revtex4-1}
\usepackage{amssymb,latexsym}
\usepackage{amsmath,amsbsy,bbm}
\usepackage{graphicx,comment}
\usepackage{epsfig,bm}
\usepackage{slashed}
\usepackage{lipsum}
\DeclareMathOperator{\tr}{tr}

\newcommand{\nn}{\nonumber}
\newcommand\beq{\begin{eqnarray}}
\newcommand\eeq{\end{eqnarray}}

\newcommand\Eq[1]{Eq.~\eqref{eq:#1}}
\newcommand\Sec[1]{Sec.~\ref{sec:#1}}
\newcommand\Ap[1]{Appendix~\ref{sec:#1}}

\newcommand{\fr}[2]{{\frac{#1}{#2}\,}}

\newcommand{\bPsi}{\bar{\Psi}}

\newcommand{\vB}{\vec{B}}
\newcommand{\vD}{\vec{D}}
\newcommand{\vE}{\vec{E}}
\newcommand{\vk}{\vec{k}}

\newcommand{\vp}{\vec{p}}

\newcommand{\vx}{\vec{x}}
\newcommand{\vsig}{\vec{\sigma}}

\newcommand{\vga}{\vec{\gamma}}

\newcommand{\calH}{\mathcal{H}}
\newcommand{\calG}{\mathcal{G}}
\newcommand{\calL}{\mathcal{L}}
\newcommand{\calE}{\mathcal{E}}
\newcommand{\calM}{\mathcal{M}}
\newcommand{\calN}{\mathcal{N}}
\newcommand{\calO}{\mathcal{O}}
\newcommand{\calS}{\mathcal{S}}
\newcommand{\calT}{\mathcal{T}}
\newcommand{\calP}{\mathcal{P}}

\begin{document}

\preprint{RBRC-1082}

\title{Reconciling the lattice background field method with nonrelativistic QED: Spinor case}

\author{Jong-Wan~Lee}
\email[]{jwlee2@ccny.cuny.edu}
\affiliation{
Department of Physics,
        The City College of New York,
        New York, NY 10031, USA}
\author{Brian~C.~Tiburzi}
\email[]{btiburzi@ccny.cuny.edu}
\affiliation{
Department of Physics,
        The City College of New York,
        New York, NY 10031, USA}
\affiliation{
Graduate School and University Center,
        The City University of New York,
        New York, NY 10016, USA}
\affiliation{
RIKEN BNL Research Center,
        Brookhaven National Laboratory,
        Upton, NY 11973, USA}
\date{\today}

\pacs{12.39.Hg, 13.40.Gp, 13.60.Fz, 14.20.Dh}

\begin{abstract}
We show that inconsistency between background field methods, 
which are relevant for lattice QCD spectroscopy, 
and effective field theory matching conditions, 
which are obtained from scattering amplitudes, 
can be resolved by augmenting nonrelativistic QED with operators related by the equations of motion.
To determine the coefficients of such operators, 
we perform the nonrelativistic expansion of QED 
for a spin-half hadron including non-minimal electromagnetic couplings. 
As an effective field theory framework could provide a valuable tool to analyze lattice QCD correlation functions
in external fields, 
we investigate whether nonrelativistic QED can be used to this end. 
We argue, 
however, 
that the most desirable approach is a hybrid one, 
which combines a relativistic hadron theory with operator selection based on nonrelativistic QED power counting. 
In this hybrid framework, 
new results are obtained for charged spin-half hadrons in uniform magnetic fields, 
including a proper treatment of Landau levels both in infinite volume and on a torus. 
\end{abstract}
\maketitle

\section{Introduction}

Theoretical understanding of low-energy hadronic properties 
is a challenging problem because of the nonperturbative nature 
of quark and gluon interactions in QCD.
Lattice gauge theory techniques provide a first principles approach to compute these properties of hadrons. 
Dramatic progress continues to be made on this front.

A standard method to determine hadronic properties is the computation of hadronic matrix elements of an
external current. 
There are a number observables, 
however, 
that require the insertion of two currents. 
For example, 
there is a long-standing discrepancy concerning the electromagnetic structure of the pion. 
Charged pion polarizabilities computed using chiral perturbation theory%
~\cite{Holstein:1990qy}
disagree with scattering experiments by a factor of 2, 
corresponding to a significance of about 
$2.5 \, \sigma$%
~\cite{Ahrens:2004mg}. 
This interesting low-energy property of pions is also relevant for high-precision physics, 
specifically the polarizability enters hadronic corrections to the muon anomalous magnetic moment%
~\cite{Engel:2012xb,Engel:2013kda,Colangelo:2014dfa}. 
The magnetic polarizability of the nucleon is another intrinsically interesting low-energy property from the point of view of chiral dynamics%
~\cite{Griesshammer:2012we}. 
Knowledge of this quantity would improve our theoretical understanding of proton structure corrections to the spectrum of muonic hydrogen%
~\cite{Hill:2011wy}, 
and allow us a better theoretical determination of the electromagnetic mass splitting between nucleons%
~\cite{WalkerLoud:2012bg}.%
\footnote{
One can compute electromagnetic mass splittings of hadrons directly from lattice QCD + QED computations, 
and considerable progress has been made recently%
~\cite{Borsanyi:2014jba}.
In this context, 
non-relativistic QED has another application, 
namely in accounting for finite volume effects from the photon, 
see~\cite{Davoudi:2014qua}.
} 
The study of hadronic polarizabilities with lattice QCD can be achieved by computing matrix elements of two electromagnetic currents. 
The computation of matrix elements of multiple currents, 
however, 
is largely beyond the reach of current lattice QCD simulations%
\footnote{
There has recently been progress on the determination of the neutral kaon mass splitting using the insertion of two weak current operators%
~\cite{Christ:2012se,Bai:2014cva}.
If the technique can be put into practice for the study of meson polarizabilities, 
however,
one must additionally confront the restriction to lattice quantized momentum.  
}

An alternative to direct computation of the Compton scattering tensor on the lattice is provided by the background field method.
This method was utilized in the early days of lattice QCD in order to obtain nucleon magnetic moments from computation of the Zeeman splitting induced by a classical magnetic field%
~\cite{Martinelli:1982cb,Bernard:1982yu}. 
The imposition of a classical electric field allows one to determine electric polarizabilities of neutral hadrons via computation of the quadratic Stark effect%
~\cite{Fiebig:1988en}.
Since these original studies, 
the background field technique continues to be refined 
and employed for the computation of other hadronic properties in lattice QCD%
~\cite{Lee:2005ds,Christensen:2004ca,Lee:2005dq,Shintani:2006xr,
Shintani:2008nt,Engelhardt:2007ub,Aubin:2008qp,Detmold:2009dx,Alexandru:2009id,
Detmold:2010ts,Alexandru:2010dx,Freeman:2012cy,Primer:2013pva,Lujan:2013qua,
Freeman:2013eta,Lujan:2014kia,Freeman:2014kka}.

To utilize the background field method, 
one must understand the behavior of hadronic correlation functions in external electromagnetic fields, 
and the relation between the extracted spectroscopic parameters with those determined from scattering experiments. 
Both can be achieved in an effective field theory framework, 
namely that of nonrelativistic QED%
~\cite{Caswell:1985ui}.
We previously showed, 
however, 
that the standard nonrelativistic QED framework leads to inconsistencies when compared with the background field method%
~\cite{Lee:2013lxa}.
For the case of a composite scalar, 
we found that nonrelativistic scalar QED must be augmented by equation-of-motion operators in order to resolve the inconsistency. 
In the present work, 
we extend our analysis to composite spin-half fermions. 
For this case, 
we also find an inconsistency between nonrelativistic QED and the background field method which can be resolved by  the inclusion of equation-of-motion operators.  
Additionally we investigate the extent to which a strict effective field theory approach based on nonrelativistic QED can be employed to analyze lattice QCD correlation functions in external fields. 
We argue, 
however, 
that the most desirable approach is a hybrid scheme based on a relativistic theory with operators chosen by their relevance in the nonrelativistic limit.


The organization of this paper is as follows. 
In \Sec{nrqed_spin_half}, 
we consider nonrelativistic QED for a charged, 
spin-half hadron. 
This theory we augment by including equation-of-motion operators, 
which are ordinarily redundant. 
Such additional operators are required, 
however,  
to resolve an inconsistency between the background field method and nonrelativistic QED matching conditions
for spin-half particles. 
After exposing this inconsistency, 
we proceed with its resolution by determining the coefficients of equation-of-motion operators. 
To obtain these coefficients, 
we perform a direct nonrelativistic expansion of the underlying relativistic theory 
with the aid of the Foldy-Wouthuysen transformation. 
The determined coefficients are shown to resolve the inconsistency between the background field method 
and nonrelativistic QED. 
In \Sec{corr_functions}, 
we investigate whether an augmented nonrelativistic QED framework is feasible for the analysis of background field correlation functions obtained from lattice QCD. 
For this investigation, 
we perform our computations in Euclidean space. 
Using the example of spin-half hadrons in an external electric field, 
we expose the difficulties involved in employing an effective field theory approach. 
Unknown hadronic parameters and error propagation are the main complications. 
The practicable alternative to a strict effective field theory approach is a hybrid approach that utilizes 
a relativistic hadron theory with operators chosen based on their relevance in the nonrelativistic limit. 
While such a relativistic theory is more complicated to work with compared to nonrelativistic QED, 
we obtain reduced theories for specific choices of uniform electric and magnetic fields. 
The derivation of these reduced theories is contained in \Ap{reduced_theory}. 
For the case of a uniform electric field, 
the reduced theory we obtain is that employed by%
~\cite{Detmold:2010ts}, 
as are the corresponding boost-projected correlation functions. 
Utilizing the velocity power counting of nonrelativistic QED, 
we obtain a useful approximation for the correlation function of charged spin-half hadrons in uniform electric fields. 
Then, we move on to the case of magnetic fields, 
where we obtain the correlation function for a charged spin-half hadron in a uniform magnetic field.  
The rather intricate form of this correlation function can be considerably simplified by isolating Landau levels. 
Using spin and parity projections, 
we show how to isolate the lowest Landau level
both in infinite volume, 
and on a torus. 
The two-point correlation functions on a torus are detailed in 
\Ap{Euclidean_torus}. 
Finally a brief summary in 
\Sec{summary} 
concludes our work, 
and outlines avenues for future investigation.

\section{Nonrelativistic QED for a spin-half hadron}
\label{sec:nrqed_spin_half}

We begin by considering the nonrelativistic QED Lagrangian for a spin-half hadron. 
The Lagrangian describes the effective theory of a composite particle coupled to electromagnetic fields. 
Typically operators in this theory are organized by inverse powers of the hadron's mass
$M$, 
and this effective theory has been previously developed to order 
$1/M^3$ 
in%
~\cite{Caswell:1985ui,Kinoshita:1995mt,Manohar:1997qy},  
and to order 
$1/M^4$ 
in%
~\cite{Heinonen:2012km,Hill:2012rh}.

\subsection{Nonrelativstic QED Lagrangian}

In formulating an effective theory, 
certain customary assumptions are made to write down simultaneously the most general and most economical Lagrangian. 
For example, 
the equations of motion are exclusively used to reduce the number of operators appearing in the effective theory. 
This reduction of operators can be rigorously justified at the level of the path integral by performing field redefinitions, 
see%
~\cite{Arzt:1993gz}, 
and references therein. 
A theory containing equations-of-motion operators and the corresponding reduced theory without such operators are equivalent provided the parameters of each theory are determined by matching $S$-matrix elements. 
Despite the complete physical equivalence of the full and reduced theories, 
their Green's functions generally differ off-shell.

Standard nonrelativistic QED needs to be augmented in order to address results from lattice QCD computations in background electromagnetic fields. 
Lattice gauge theory methods give one access to hadronic Green's functions, 
and typically the mixed momentum-time representation of the Green's function restricts one to on-shell hadronic states. 
We previously detailed that lattice QCD correlation functions for scalar hadrons in time-dependent external fields, 
however, 
require that certain equation-of-motion operators be retained%
~\cite{Lee:2013lxa}.
This peculiarity is tied to the lack of an on-shell condition. 
Fortunately one can still study the behavior of the scalar hadron's Green's function to extract physical properties,
but care is required.

Our present goal is to reconcile nonrelativistic QED with the background field method for the case of spinor hadrons. 
The first step is to retain certain equation-of-motion operators in nonrelativistic QED. 
By imposing Hermiticity, and invariance under $P$, $T$, and gauge transformations, 
we find the augmented nonrelativistic QED Lagrangian up to 
$O(M^{-3})$
has the form
\begin{widetext}
\beq
\calL&=&\psi^\dagger\left[iD_0+c_2\fr{\vec{D}^2}{2M}+c_4\fr{\vec{D}^4}{8M^3}
+c_F\fr{\vec{\sigma}\cdot\vec{B}}{2M}
+c_D\fr{[\vec{\nabla}\cdot\vec{E}]}{8M^2}
+ic_S\fr{\vec{\sigma}\cdot(\vec{D}\times\vec{E}-\vec{E}\times\vec{D})}{8M^2}
+c_{W_1} \fr{\{\vec{D}^2,\vec{\sigma}\cdot\vec{B}\}}{8M^3}
\right. \nn \\
&&
\left.
-c_{W_2}\fr{D^i \vec{\sigma}\cdot\vec{B} D^i}{4M^3}
+c_{p'p}\fr{ \big\{ \vec{D}\cdot\vec{B}, \vec{\sigma}\cdot\vec{D} \big\}
}{8M^3}
+c_{A_1}\fr{\vec{B}^2-\vec{E}^2}{8M^3}
-c_{A_2}\fr{\vec{E}^2}{16M^3}
+ic_M\fr{\{D^i,[\vec{\nabla}\times\vec{B}]^i\}}{8M^3}
\right. \nn \\
&&
\left.
+c_{X_0}\fr{[iD_0,\vec{D}\cdot\vec{E}+\vec{E}\cdot\vec{D}]}{8M^3}
+c_{X'_0}\fr{ \big[ D_0, [ D_0, \vec{\sigma}\cdot \vB 
] \big]}{8M^3}
\right]\psi,
\label{eq:nrqed}
\eeq
\end{widetext}
where $D_0=\partial_0-iZA_0$ and $D^i=\nabla^i-iZA^i$. 
The electric and magnetic fields $\vec{E}$ and $\vec{B}$ are given by 
standard expressions, $\vec{E}=-\partial_0\vec{A}-\vec{\nabla}A^0$ and 
$\vec{B}=\vec{\nabla}\times\vec{A}$, respectively. 
Note that we have adopted the convention that bracketed derivatives only 
act on quantities appearing inside the square brackets.

The last two operators appearing in the above Lagrangian are new, 
and are ordinarily redundant in light of the equations of motion. 
These operators, 
however, 
can explicitly modify the time dependence of Green's functions 
in certain time-dependent background gauge fields.
\footnote{
Notice that there are more equation-of-motion operators, 
such as 
$\psi^\dagger (iD_0)^n \psi$,
for 
$n>1$, 
and 
$\vE^2 \psi^\dagger iD_0 \psi$. 
The former operators modify Green's functions by 
singular terms involving time derivatives of delta functions, 
$(i D_0)^{n-1}\delta(t'-t)$, 
while the latter operator modifies the time dependence of the Green's function
only in a time-dependent electric field. 
We need not consider the former operators, 
and the latter will not appear given the added assumption of a canonical form for the kinetic term. 
Indeed, 
with the explicit reduction of the relativistic Lagrangian below, 
we do not find operators of the latter type. 
}
For on-shell processes involving the
$\psi$
particle, 
these two operators 
can be eliminated from the Lagrangian using the equations of motion
\beq
\psi^\dagger\fr{[iD_0,\vD\cdot\vE+\vE\cdot\vD]}{8M^3}\psi
&=&-c_2\psi^\dagger\fr{[\vD^2,\vD\cdot\vE+\vE\cdot\vD]}{16M^4}\psi, \nn\\
 \psi^\dagger\fr{\big[ D_0, [ D_0, \vec{\sigma}\cdot \vB ] \big]}{8M^3}\psi
&=& - c_2^2 \psi^\dagger\fr{\big[\vD^2, [ \vD^2, \vsig\cdot \vB ]  \big]}{32 M^5}\psi,
\nn \\ 
\label{eq:eom_ops}
\eeq
and what remains is the standard nonrelativistic QED Lagrangian. 
In particular, 
the first operator gives rise to the 
$c_{X_1}$
term of the 
$O(M^{-4})$
Lagrangian%
~\cite{Hill:2012rh}, 
while 
the second operator gives rise to an operator of yet higher order. 
Off-shell, 
however, 
these operators will modify Green's functions and need to be accounted for to describe lattice QCD correlation functions
in external fields, 
for example. 
We stress that there is absolutely no modification necessary to standard nonrelativistic QED matching conditions that utilize scattering amplitudes. 
Our considerations extend beyond the standard matching to the Green's functions themselves; 
and, 
for this reason,
we must retain equation-of-motion operators.

To demonstrate their relevance, 
we show how the absence of equation-of-motion operators leads to physically unreasonable results. 
In particular, 
we find an incorrect relation between static hadron properties and scattering observables.
This is the spin-half generalization of the example elaborated upon in%
~\cite{Lee:2013lxa}. 
Taking the zero-momentum limit of the standard nonrelativistic QED Lagrangian
for the case of a uniform external electric field, 
we find the energy is shifted by%
~\cite{Hill:2012rh}
\beq
-\delta E(\vE)=-\left(c_{A_1}+\fr{1}{2}c_{A_2}\right)
\fr{\vE^2}{8M^3}+\cdots,
\label{eq:energy_shift_const_E}
\eeq
where terms of higher order than 
$M^{-3}$ 
have been dropped. 
Notice that 
this energy shift arises only from interactions with real photons. 
On the other hand, 
the nonrelativistic QED matching conditions, 
which are 
obtained from one- and two-photon scattering amplitudes%
~\cite{Hill:2012rh}
can be used to rewrite the linear combination of low-energy constants in terms of physically measurable observables, 
namely
\beq
c_{A_1}+\fr{1}{2}c_{A_2}=Z^2 + \kappa^2-16\pi M^3 \alpha_E+\fr{4}{3} Z M^2\langle r^2_E \rangle , 
\quad \label{eq:NRQED_matching_const_E}
\eeq
where 
$\kappa$, 
$\alpha_E$, 
and 
$\langle r_E^2 \rangle$ 
are the anomalous magnetic moment, 
electric polarizability, and electric charge radius, 
respectively. 
Because the contribution from the charge radius can only be resolved from a virtual photon, 
we are confronted with a physical contradiction between the results of 
Eqs.~(\ref{eq:energy_shift_const_E}) and (\ref{eq:NRQED_matching_const_E}). 
As we will see below, 
this contradiction is resolved by retaining operators related by equations of motion, 
in particular the term with coefficient 
$c_{X_0}$. 
The coefficients of equation-of-motion operators must be determined, 
however, 
and this we achieve through the explicit expansion of relativistic QED in the nonrelativistic limit.

\subsection{Foldy-Wouthuysen transformation}
\label{sec:FW}

A long time ago, 
Foldy and Wouthuysen performed a canonical transformation on the Dirac Hamiltonian 
that leads to a new representation of the Dirac theory in which positive and 
negative energy states are separately represented by two-component wave-functions%
~\cite{Foldy:1949wa}. 
In this new representation, 
the nonrelativistic expansion is manifest because 
the positive energy states are decoupled from the negative energy states. 
In this section, 
we briefly review the Foldy-Wouthuysen (FW) transformation using a Lagrangian framework.

We first consider the simple example of a free Dirac fermion
\beq
\calL&=&\bar{\Psi}[i\slashed{\partial}-M]\Psi.
\label{eq:free_Dirac}
\eeq
In Minkowski space-time,  
we 
adopt the metric 
$\eta^{\mu\nu}=\delta^{\mu\nu}\{-1,1,1,1\}$,  
with Dirac matrices satisfying the anti-commutation relations 
\beq
\{\gamma^\mu,\gamma^\nu\}=-2\eta^{\mu\nu}I_4,
\eeq 
where 
$I_4$ 
is the 
$4\times4$ identity matrix. 
Any Dirac matrix can be classified as even or odd based on its commutation relations with 
$\gamma^0$. 
The even operators,
$\calO_E$, 
satisfy
\beq
\left[ \calO_E, \gamma^0 \right] = 0
,\eeq
while odd operators,
$\calO_O$,
satisfy
\beq
\left\{ \calO_O, \gamma^0 \right\} = 0
.\eeq
In the Dirac basis, 
even operators do not mix the upper and lower two-components of the Dirac fermion, 
while odd operators connect the upper and lower components. 
In order to separate the positive and negative energy states, 
we need to systematically eliminate the odd operators. 
To this end, 
we consider the following unitary transformation
\beq
\Psi'=e^{iS}\Psi,
\eeq
where $S$ is a Hermitian operator which generally can be time dependent.

For a free Dirac fermion,  
it is convenient to work in momentum space in order to show how the transformation proceeds. 
Taking  
\Eq{free_Dirac}
in momentum space, 
we define the box operator by
$\calL = \Psi^\dagger \Box \Psi$, 
with
\beq
\Box= i\partial_0-\gamma^0\vec{\gamma}\cdot\vec{k}-\gamma^0 M.
\eeq
We want to transform away the
$\gamma^i$'s, 
which are odd operators,  
and thus take the form of 
$S$ 
to be
\beq
S=-\fr{i}{2M}\vec{\gamma}\cdot\vec{k} \, \,\, \omega\Big(\fr{k}{M}\Big),
\eeq
with 
$k = | \vk |$.
In order to determine the function 
$\omega$ 
such that the transformed operator, 
$\Box' = e^{iS}\Box e^{-iS}$, 
is free of odd operators, 
we explicitly evaluate 
$\Box'$
in this simple example
\beq
\Box'
&=&
i\partial_0-\gamma^0
\left[
M\cos (k \omega/M)+k \sin ( k \omega/M)
\right]
\nn \\
&&
-\gamma^0\fr{\vec{\gamma}\cdot\vec{k}}{k}
\left[
k \cos ( k \omega/M)-M\sin (k\omega/M)
\right].
\eeq
Demanding the last term vanishes leads us to the functional form
$\omega(x)=\frac{1}{x} \tan^{-1} x$, 
and 
\beq
\Box'=i\partial_0-\gamma^0 E_{k},
\label{eq:positive_free_action}
\eeq
with 
$E_{k} = \sqrt{M^2+k^2}$.
As expected, 
in the new representation 
positive and negative energy states can be cleanly separated by simply projecting according to eigenvalues of the
$\gamma^0$
matrix
\beq
\psi_\pm&=&\calP_\pm \Psi',
\eeq
where
$\calP_\pm = \fr{1}{2} \left( 1 \pm \gamma^0\right)$.
This clean separation is not possible in the original representation, 
\emph{viz.}
\beq
\Psi_\pm&=&e^{-iS}\calP_\pm e^{iS}\Psi.
\eeq

Given the relativistic Green's function
\begin{equation} 
G(x)=\langle 0 | T\left\{ \Psi(x)\bar{\Psi}(0)  \right\} | 0 \rangle
,\end{equation} 
the positive energy Green's function can be derived by utilizing the FW transformation, 
namely
\beq
G_+(x)
&\equiv&
\theta (x^0)
\langle 0 | \psi_+(x)\psi^\dagger_+(0)
| 0 \rangle \nn \\
&=&
\theta (x^0)
\calP_+ e^{iS}G(x)\gamma^0 e^{-iS}
\calP_+
.\label{eq:positive_G}
\eeq
For a free Dirac fermion, 
the mixed momentum-time relativistic Green's function is given by
\beq
G(x^0, \vec{k})
&=&
\frac{e^{ - i  E_{k} | x^0| }}{2 E_{k}}
\Bigg[ 
\theta(x^0)
\left( 
\gamma^0 E_{k}  
- 
\vga \cdot \vk
+ 
M
\right)
\notag \\
&&
\phantom{pacing}
- 
\theta(-x^0)
\left(
\gamma^0 E_{k} 
+ 
\vga \cdot \vk
-
M
\right)
\Bigg]
.\quad \, \eeq
Using \Eq{positive_G}, 
we arrive at the positive energy Green's function
\beq
G_+(x^0, \vk)
=
\theta(x^0)
\calP_+
e^{ - i  E_{k}  x^0 }
,\eeq
which is precisely the same Green's function we obtain directly from \Eq{positive_free_action}. 
A salient feature of the FW transformation is that it properly incorporates the standard nonrelativistic normalization of states.

Having considered a free Dirac fermion,  
let us move on to the interacting case. 
The procedure is the same as the FW transformation of the free Dirac fermion, 
except for determination of the function 
$\omega$. 
Due to the presence of interactions, 
the Lagrangian can now include more than one odd operator, 
and it becomes very difficult or even impossible to find a 
closed-form expression for the function
$\omega$. 
One possible way to find the new representation is 
to eliminate the odd operators order-by-order in 
$1/M$, 
from which we will naturally arrive at the nonrelativistic QED Lagrangian.
An additional complication arises due to the non-renormalizability of the effective hadronic theory. 
Higher-derivative operators are present in the effective theory;
thus, 
in order to have manifest power counting in 
$1/M$, 
we must perform the phase transformation
$\Psi \longrightarrow e^{ - i M x_0 } \Psi$, 
followed by the FW transformation.
In general, 
one may write the interacting Lagrange density after these transformations in the form  
\beq
\label{eq:Dirac_action}
\Box
=
i\partial_0
+ 
( 1- \gamma^0)
M
-
\calO_O
-
\calO,
\eeq
where 
$\calO_O$ 
is an odd operator, 
and 
$\calO$
contains both even and odd operators. 
In what follows, 
we will need multiple iterations of the procedure, 
and so we take 
$\calO_O$ 
to include odd operators at a given order, 
while 
$\calO$ 
includes even operators at all orders, 
and only odd operators of higher-order. 
The FW transformation of the operator 
$\Box$ 
is given by
\beq
\Box'&=&e^{iS}\Box e^{-iS}\nn \\
&=&\Box+[iS,\Box]+\fr{1}{2!}[iS,[iS,\Box]]+\fr{1}{3!}[iS,[iS,[iS,\Box]]]
\nn \\
&&+\cdots,
\eeq
where 
$S$
is chosen to be
$S=-i\gamma^0 \calO_O/ 2 M$. 
With this choice for 
$S$, 
one can show that the transformed operator takes the form
\beq
\label{eq:FW_projection}
\Box'&=&i\partial_0
+ 
( 1- \gamma^0)
M
- 
\calO
-
\fr{[\gamma^0\calO_O,\calO-i\partial_0]}{2M}
-
\fr{\gamma^0 \calO_O^2}{2M}
\nn \\
&&-\fr{[\gamma^0\calO_O,[\gamma^0\calO_O,\calO-i\partial_0]]}{8M^2}
+\fr{\calO_O^3}{3M^2}
\nn \\
&&
-\fr{[\gamma^0\calO_O,[\gamma^0\calO_O,[\gamma^0\calO_O,\calO-i\partial_0]]]}{48M^3}
+\fr{\gamma^0 \calO_O^4}{8M^3}+\cdots.
\notag \\
\eeq
As desired, the odd operator 
$\calO_O$
at the given order has been eliminated, 
while new even and odd operators appear at higher orders in the 
$1/M$ 
expansion. 
Starting from the lowest odd operator remaining, we iteratively perform the
transformation until we obtain the nonrelativistic action free from odd operators to the desired order in 
$1/M$.

\subsection{Relativistic QED Lagrangian}

Using the FW transformation, 
we can derive the nonrelativistic QED action for a charged spin-half hadron 
from the relativistic action 
which includes nonminimal couplings to electromagnetic fields. 
From this explicit expansion, 
we directly obtain the nonrelativistic QED matching conditions, 
including those for operators related by equations of motion.

To write down the effective action for 
$\Psi$, 
we enforce the usual 
$C$, 
$P$, 
and 
$T$ 
invariance in addition to Lorentz and gauge invariance. 
The operators of this action are largely organized in powers of the inverse hadron mass 
$M$, 
however, 
terms with time derivatives acting on the relativistic field 
$\Psi$
field are ultimately promoted to lower order in the nonrelativistic limit. 
As a result,  
we cannot write down all possible operators;
rather, 
we write down all operators that are relevant to nonrelativistic QED through 
$O(M^{-3})$.
\footnote{
A natural question to ask is whether one must include equation-of-motion operators in the relativistic theory as well. 
For example, 
higher dimensional operators involving 
$\bPsi \cdots \fr{(i\slashed{D})^n}{M^{n-1}}\Psi$ 
are all the same order in the nonrelativistic limit. 
Such operators, 
however, 
can be eliminated using the equations of motion even with external electromagnetic fields, 
because the required field redefinitions modify the Green's function
$G(x',x)$ 
only by singular terms at 
$x'=x$, 
namely 
$( i \slashed{D})^{n-1}\delta(x'-x)$. 
Furthermore, we exclude operators of the form 
$F^2\bPsi (i\slashed{D})\Psi$, 
because they modify the Green's function by an overall constant in 
uniform electromagnetic fields. 
Both classes of equation-of-motion operators can be ignored with the added assumption of a canonical form
for the kinetic term in the Lagrangian. 
}
With this limit in mind, 
we have the following relativistic Lagrange density for a charged composite spin-half hadron
\beq
\calL&=&\bPsi\Big[i\slashed{D}-M
+\fr{\kappa}{4M}\sigma_{\mu\nu}F^{\mu\nu}
-\fr{C_1}{M^2}\gamma_\mu[D_\nu, F^{\mu\nu}]
\nn \\
&&
+
\fr{C_2}{M^3}
\sigma_{\mu\nu}[D_\rho,[D^\mu,F^{\nu\rho}]]
+
\fr{C_3}{M^3}
F^{\mu\nu}F_{\mu\nu}
\Big]\Psi
\nn \\
&&
+\fr{iC_4}{M^4}\bPsi\gamma_\mu D_\nu\Psi
T^{\mu\nu}
-\fr{C_5}{M^5}
D_\mu\bPsi D_\nu\Psi
T^{\mu\nu}
\nn \\
&& -\fr{C_6}{M^5}
D_\rho\bPsi D^\rho\Psi
F^{\mu\nu}F_{\mu\nu},
\label{eq:relqed}
\eeq
where the gauge covariant derivative is
$D_\mu=\partial_\mu-iZA_\mu$, 
and the electromagnetic field strength is given by
$F^{\mu\nu}=\partial^\mu A^\nu-\partial^\nu A^\mu$. 
We define the electromagnetic stress-energy tensor as 
$T^{\mu\nu}
=
F^{\rho\{\mu}F^{\nu\}} {}_{\rho}$, 
where the curly braces denote symmetrization and trace subtraction
\beq
\calO^{\left\{\mu\nu\right\}}=\frac{1}{2}\left(
\calO^{\mu\nu}+\calO^{\nu\mu}-\fr{1}{2}\eta^{\mu\nu}\calO_\alpha^{~\alpha}
\right).
\eeq
It is useful to recall that commutators of two covariant derivatives can be expressed in terms of electric and magnetic fields, 
namely
$[D^i,D_0]=-iZE^i$ 
and 
$[D^i,D^j]=-iZ\epsilon^{ijk}B^k$.
The matrix
$\sigma_{\mu\nu}$
has the usual definition,
$\sigma_{\mu \nu}=\fr{i}{2}[\gamma_\mu,\gamma_\nu]$.

Notice that the final three operators appearing in \Eq{relqed} will be relevant to 
$O(M^{-3})$ 
in the nonrelativistic limit because they contain time derivatives acting on the massive hadron field. 
In addition to the mass, 
the relativistic Lagrangian depends on seven coefficients, 
which can be determined by computing the amplitudes for one- and two-photon processes
and comparing with experiment. 
Values for these coefficients can be determined by using physical quantities.
Specifically we will see these quantities are
the magnetic moment
$\mu$, 
the charge and magnetic radii,
$<r_E^2>$
and
$<r^2_M>$, 
as well as the electric and magnetic polarizabilities, 
$\alpha_E$
and
$\beta_M$. 
These five physical quantities leave two linear combinations of coefficients undetermined, 
however, 
these linear combinations cannot be resolved at 
$O(M^{-3})$. 
Carrying out the matching, 
we find that 
$\kappa$ 
is the anomalous magnetic moment, 
$\kappa = \mu - Z$, 
while 
$C_1$
and 
$C_2$
are simply related to the Dirac and Pauli radii, 
respectively. 
These relations can be rewritten in terms of the conventionally defined charge and magnetic radii
\beq
<r_E^2>
&=&
\frac{3}{2 M^2} 
\left( 4 C_1 + \kappa \right)
,\nn \\
<r_M^2>
&=&
\frac{6}{M^2}
\left(
C_1
- 2 C_2
\right)
\label{eq:radii}
.\eeq
Lastly from the amplitude for the real Compton scattering process, 
we find
\beq
\pi M^3
\alpha_E
&=&
- C_3- C_6+\fr{1}{4}(C_4+C_5)
,\nn \\  
\pi M^3
\beta_M
&=&
\phantom{-}
C_3+C_6+\fr{1}{4}(C_4+C_5)
\label{eq:pols}
.\eeq

\subsection{Nonrelativistic limit and matching}

To take the nonrelativistic limit, 
we perform a series of FW transformations. 
After the phase transformation, 
$\Psi \to e^{ - i M x_0 } \Psi$,
the relativistic Lagrangian in \Eq{relqed} has the form of 
\Eq{Dirac_action}, 
where we identify the leading-order odd operator as
\beq
\calO_O&=&-i\gamma^0\vec{\gamma}\cdot\vec{D},
\eeq
with all even operators and the remaining odd operators appearing in 
$\calO$. 
To make powers of the inverse hadron mass manifest, 
we write the expansion in the form
\beq
\calO
&=&
\calO_E^{(0)}
+
\fr{1}{M}(\calO^{(1)}_O+\calO^{(1)}_E)
\nn \\
&&
+\fr{1}{M^2}(\calO^{(2)}_O+\calO^{(2)}_E)
+\fr{1}{M^3}\calO^{(3)}_E
+\cdots.
\eeq
The omitted higher-order terms include odd operators at 
$O(M^{-3})$. 
We need not consider these operators, 
however, 
because they are eliminated by the FW transformation, 
and generate even operators beginning at 
$O(M^{-4})$ 
in the nonrelativistic limit. 
In the expansion of 
$\calO$, 
the even operators at each order are given by 
\beq
\calO^{(0)}_E
&=&
- Z  A_0
,\nn \\
\calO^{(1)}_E&=&- \kappa \gamma^0\vec{\Sigma}\cdot\vec{B}
,\nn \\
\calO^{(2)}_E&=&
-C_1[D^i,E^i],
\eeq
and
\beq
\calO^{(3)}_E
&=&
2 \left(C_3+C_6\right)\gamma^0 (\vec{B}^2 - \vec{E}^2)
+
\fr{C_5}{2} \gamma^0 (\vec{B}^2 + \vec{E}^2 )
\nn \\
&&
+
\fr{C_4}{2} ( \vec{B}^2 + \vec{E}^2 )
+
2
C_2\gamma^0\left[D_0, [ D_0, \vec{\Sigma} \cdot \vB] \right]
\nn \\
&&
- 
2 C_2\gamma^0\{\vD^2,\vec{\Sigma} \cdot\vB\}
+
4 C_2 \gamma^0 D^i \vec{\Sigma} \cdot\vB D^i.
\eeq
The odd operators beyond leading order have the form
\beq
\calO^{(1)}_O&=&\fr{i\kappa}{2}\vec{\gamma}\cdot\vec{E}
,\nn \\
\calO^{(2)}_O&=&
-C_1[D_0,\gamma^0\vec{\gamma}\cdot\vec{E}]
+C_1[\gamma^0(\vec{\gamma}\times\vec{D})^k,B^k].
\label{eq:odds}
\eeq
To arrive at these operators, 
we recall that the Dirac matrices satisfy the relation
$\gamma^i\gamma^j=-(\delta^{ij} I_4 +2 i\epsilon^{ijk}\Sigma^k)$, 
which is written in terms of spin operators obeying the Lie algebra
$\left[ \Sigma^i, \Sigma^j \right] =  i \epsilon^{ijk} \Sigma^k$, 
along with the property
$\left\{ \Sigma^i, \Sigma^j \right\} = \frac{1}{2} \delta^{ij} I_4$.

To eliminate the odd operators in \Eq{odds} from the action, 
we take the following three FW transformations: 
\beq
S_1=-\fr{\vga\cdot\vD}{2M},
\label{eq:S1}
\eeq 
to eliminate
$\calO_O$,
then
\beq
S_2=
(Z+\kappa)
\fr{\gamma^0\vga\cdot\vE}{4M^2},
\label{eq:S2}
\eeq 
to eliminate 
$\calO_O^{(1)}$, 
and finally
\beq
S_3
&=&
\fr{(\vga\cdot\vD)^3}{6M^3}
+
(Z+\kappa+4C_1)
\fr{[iD_0,\vga\cdot\vE]}{8M^3}\nn \\
&&
-
iC_1
\fr{[(\vga\times\vD)^i,B^i]}{2M^3}
-\kappa\fr{\{\vga\cdot\vD,\vec{\Sigma}\cdot\vB\}}{4M^3},
\label{eq:S3}
\quad \eeq
to eliminate 
$\calO_O^{(2)}$, 
as well as the odd operators produced after accounting for the 
$S_1$
and
$S_2$
transformations. 
After this series of three FW transformations, 
the action density in the new representation is given by
$\calL=\Psi'^\dagger \Box' \Psi'$, 
with
\beq
\Box'
=
e^{i S_3}
e^{i S_2}
e^{i S_1}
\Box
e^{ - i S_1}
e^{- i S_2}
e^{- i S_3}
.\eeq
Carrying out these transformations
using \Eq{FW_projection}, 
we obtain 
\beq
\Box' 
= 
i D_0 
+ 
(1 - \gamma^0) 
M
+ 
\frac{\calO_E^{\prime(1)}}{M}
+ 
\frac{\calO_E^{\prime(2)}}{M^2}
+ 
\frac{\calO_E^{\prime(3)}}{M^3}
,
\label{eq:FWBox}
\eeq
which is free from odd operators.
The even operators at 
$O(M^{-1})$ 
are given by 
\beq
\calO_E^{\prime(1)}
=
\frac{1}{2}
\gamma^0
\vec{D}^2
+
(Z+\kappa)\gamma^0
\vec{\Sigma}\cdot\vec{B}
,\eeq
while those at 
$O(M^{-2})$
are
\beq
\calO_E^{\prime(2)}
&=&
\frac{1}{8}
(Z+2\kappa+8C_1)
\vec{\nabla}\cdot\vec{E}
\nn \\
&&+
\frac{i}{4}
(Z+2\kappa)
\vec{\Sigma}\cdot(\vec{D}\times\vec{E}-\vec{E}\times\vec{D})
,\eeq
and finally those at 
$O(M^{-3})$
are found to be
\begin{widetext}
\beq
\calO_E^{\prime(3)}
&=& 
\frac{1}{8}
\gamma^0
\vec{D}^4
+
\frac{1}{8}
(2Z+\kappa+8C_1-16C_2)\gamma^0 
\{\vec{D}^2,\vec{\Sigma}\cdot\vec{B}\}
-
\frac{1}{4}
(\kappa+8C_1-16C_2)
D^i \vec{\Sigma}\cdot\vec{B} D^i
+
\frac{1}{4}
\kappa\gamma^0 
\big\{
\vec{D}\cdot\vec{B}, \vec{\Sigma}\cdot\vec{D} 
\big\}
\nn \\
&&
+
\frac{1}{2}
C_4
\big(
\vB^2+\vE^2
\big)
+
\frac{1}{8}
(Z^2+16C_3+4C_5+16C_6)\gamma^0
\big(
\vec{B}^2-\vec{E}^2
\big)
-
\frac{1}{8}
(\kappa^2+2Z\kappa+8ZC_1-8C_5)\gamma^0 \vec{E}^2\nn \\
&&
+
\frac{i}{16}
(\kappa+8C_1)\gamma^0
\{D^i,(\vec{\nabla}\times\vec{B})^i\}
-
\frac{i}{2}
C_1\gamma^0
[D_0,\vec{D}\cdot\vec{E}+\vec{E}\cdot\vec{D}]
-
(C_1-2C_2)\gamma^0
\big[ D_0, [ D_0, \vec{\Sigma}\cdot \vB ] \big]
.\label{eq:new_relqed}
\eeq
\end{widetext}

To produce the nonrelativistic QED Lagrangian from \Eq{FWBox}, 
we write the relativistic action in terms of the positive- and negative-energy projected fields, 
$\psi_\pm$, 
and finally integrate out the field 
$\psi_-$, 
which is trivially accomplished due to the FW transformation.  
For simplicity, 
we write the remaining field
$\psi_+$
as
$\psi$, 
and arrive at the nonrelativistic QED Lagrangian in \Eq{nrqed} provided the following matching conditions are met. 
At 
$O(M^{-1})$, 
we require that the coefficients satisfy
\beq
c_2=1, 
\quad 
\text{and}
\quad
c_F
= 
Z + \kappa
= 
\mu
.\eeq 
At the next order, 
$O(M^{-2})$, 
we find that matching requires the relations
\beq
c_D=Z+2\kappa+8C_1, 
\quad
\text{and}
\quad
c_S
= 
Z + 2 \kappa
,\eeq
where the former relation can be rewritten in the form
$c_D = Z + \frac{4}{3} M^2 <r_E^2>$, 
using the identification of the charge radius from \Eq{radii}.

At $O(M^{-3})$, 
there are coefficients whose values are exactly fixed by those of the lower-order coefficients, 
namely
\beq
c_4=1,
\quad
c_{p'p}=\kappa,
\quad \text{and} \quad
c_M= \frac{1}{2} ( c_D - c_F)
.\eeq
The remaining coefficients of the standard nonrelativistic QED Lagrangian are given by the matching conditions
\beq
c_{W_1}
&=&
Z+c_{W_2}
=
Z
+
\fr{\kappa}{2}
+
4 ( C_1- 2 C_2),
\nn \\
c_{A_1}
&=&
Z^2
+ 
16 
\left[
C_3
+ 
C_6 
+ 
\frac{1}{4} 
\left( 
C_4
+ 
C_5
\right)
\right],
\nn \\
c_{A_2}
&=&
2\kappa^2
+
4Z\kappa
+
16 
\left(
Z C_1
- 
C_4
-
C_5
\right)
.\eeq
The first of these relations can be rewritten in terms of the magnetic radius, 
$c_{W_1} = Z + \frac{\kappa}{2} + \frac{2}{3} M^2 < r_M^2 >$,
using \Eq{radii}, 
while the second of these relations can be rewritten in terms of the magnetic polarizability, 
$c_{A_1} = Z^2 + 16 \pi M^3 \beta_M$,
with the help of \Eq{pols}. 
Forming the linear combination 
$c_{A_1} + \frac{1}{2} c_{A_2}$, 
and taking into account the expressions for polarizabilities
\footnote{
It is curious to note that in 
$\calO^{(3)}_E$, 
the 
$C_4$ 
term has the spin structure
$\calP_+ + \calP_-$,
which is different than the other 
$\vE^2$ 
and 
$\vB^2$ 
operators, 
which have the structure 
$\calP_+ - \calP_-$. 
While this would seem to imply that the electric and magnetic polarizabilities of a spin-half anti-hadron 
are different from those of a spin-half hadron, 
all of the operators in 
\Eq{relqed} are invariant under 
$C$, 
$P$, 
and 
$T$. 
To resolve this issue, 
we note that to arrive at an effective theory for the anti-hadron, 
we must employ the opposite phase transformation, 
$\Psi \to e^{ i M x_0} \Psi$. 
Because the
$C_4$
term has an odd number of derivatives, 
its leading-order contribution to the anti-hadron theory will be reversed, 
and the polarizabilities for the anti-hadron are hence the same as for the hadron. 
} 
in terms of coefficients of the relativistic theory, 
we arrive at 
\Eq{NRQED_matching_const_E}.
Thus when all of the nonrelativistic QED coefficients are expressed in terms of physical quantities, 
the matching conditions agree with those deduced in%
~\cite{Hill:2012rh}.

Two further matching conditions arise at 
$O(M^{-3})$. 
These conditions determine values for the coefficients of equation-of-motion operators, 
namely
\beq
c_{X_0}
&=&
- 4 C_1 
= 
\kappa - \frac{2}{3} M^2 < r_E^2>,
\nn \\
c_{X'_0}
&=&
- 4 (C_1 - 2 C_2)
=
- \frac{2}{3} M^2 < r_M^2>
\label{eq:eom}
.\eeq
Notice these coefficients are non-vanishing only for composite particles. 
In uniform electric fields, 
we must take into account the equation-of-motion operators in \Eq{nrqed}. 
In particular, 
the 
$c_{X_0}$ 
term makes a non-vanishing contribution to the energy shift, 
which we can now determine given the values of coefficients obtained in \Eq{eom}. 
We find the energy shift in a uniform electric field takes the form
\beq
\delta E(\vE)=-\fr{1}{2}\left(4\pi\alpha_E-\fr{\mu^2}{4M^3}\right)\vec{E}^2,
\label{eq:Eshift}
\eeq
where,
compared to
\Eq{energy_shift_const_E}, 
the contribution from the charge radius, 
$<r_E^2>$,
exactly cancels due to inclusion of the equation-of-motion operator. 
Therefore, 
we conclude that there are no inconsistencies between the background field method 
and nonrelativistic QED matching provided equation-of-motion operators are retained. 
As a result, 
hadronic parameters, 
such as the electric and magnetic polarizabilities determined from 
Compton scattering, 
can be extracted from lattice QCD calculations in background fields. 
This requires knowledge of the hadronic correlation functions, 
to which we now turn.

\section{Euclidean correlation functions}
\label{sec:corr_functions}

An essential ingredient to the background field method is the determination of correlation functions using an effective hadronic theory, 
and the comparison of these predictions with the behavior of correlation functions computed in lattice QCD. 
Because parameters of the effective theory enter scattering amplitudes, 
one can determine physical observables from knowledge of the Green's functions, 
specifically their dependence on the strength of external fields. 
In the effective hadronic theory, 
this behavior can be determined from the nonrelativistic QED action in \Eq{nrqed}, 
as well as the relativistic theory in \Eq{relqed}. 
In terms of matching, 
we have shown these two approaches are consistent, 
however,
they have their own intrinsic difficulties. 
The former makes direct comparison with lattice QCD correlators rather complicated. 
On the other hand, 
the latter lacks manifest power counting which complicates the enumeration of operators.   
In this section, 
we expound on these difficulties and compute Euclidean correlation functions 
for specific external fields, 
namely uniform electric and magnetic fields.
\footnote{
Notice that throughout this section all operators are defined in Euclidean space-time, 
where the metric is given by $\eta_E^{\mu\nu}= \delta^{\mu \nu}$. 
The conversion from Minkowski space-time 
$x^\mu=( t,\vx )$ 
to Euclidean space-time 
$x_E^\mu=(\vx,\tau)$ 
is achieved by the Wick rotation 
$t = - i \tau$. 
The Euclidean Dirac matrices are specified by 
$\gamma^4_E=\gamma^0$ 
and 
$\gamma_E^i=-i\gamma^i$, 
and satisfy the anti-commutation relation 
$\{\gamma^\mu_E,\gamma^\nu_E\}=2\eta_E^{\mu\nu}$. 
We omit the subscript 
$E$ 
for convenience in the main text. 
In the case of a uniform electric fields, 
the Euclidean space formulation avoids the Schwinger mechanism. 
To make the Euclidean formulation clear, 
we write the Euclidean electric field as 
$\vec{\calE}$, 
which is related to the Minkowski electric field,
$\vE$, 
by the analytic continuation
$\vec{\calE} = - i \vE$.  
}

\subsection{Difficulty with nonrelativistic correlation functions}
\label{sec:corr_nonrel}

Nonrelativistic QED provides an ideal framework to compute single hadron propagators at low energy 
because of the relatively simple form of interactions, 
and the manifest power counting. 
For applications to external fields, 
we use velocity power counting, 
where quantities are expanded in powers of the small velocity 
$v$. 
Spatial and temporal derivatives count differently in the velocity expansion, 
$\vec{D} \sim v$ 
and
$D_4 \sim v^2$, 
respectively. 
As a result, 
external electric and magnetic fields count as 
$\vec{\calE} \sim v^3$
and
$\vB \sim v^3$. 
The electric and magnetic polarizabilities both enter the effective theory at
$O(v^6)$, 
and thus require determination of propagators including all terms to the same order.

To exemplify the difficulties in using nonrelativistic correlators,  
we determine the nonrelativistic QED propagator in an external electric field. 
We consider the case of a uniform electric field along $x_3$-direction 
specified by the vector potential, $
\vec{A}=-\calE \tau \hat{x}_3$. 
Notice that for charged hadrons, 
the correlation functions are generally gauge variant. 
In this particular gauge, 
we can dramatically simplify the correlator by projecting onto vanishing three-momentum. 
According to velocity power counting,  
we have the Euclidean nonrelativistic QED action density to 
$v^6$ 
order
\footnote{
Compared to the nonrelativistic QED action density given in \Eq{nrqed}, 
we add one more term at 
$O(M^{-5})$, 
which is the next-order relativistic correction to the kinetic energy. 
This term is included in the action because it contributes at 
$O(v^6)$, 
just as the polarizabilities. 
}, 
\beq
\calL&=&\psi^\dagger_{\vp=0}\left[
\fr{\partial}{\partial \tau}+\fr{(Z\calE \tau)^2}{2M}
-\frac{(Z\calE \tau)^4}{8M^3}
\right.
\nn \\
&&\left.
+\frac{(Z\calE \tau)^6}{16M^5}
+c_{\textrm{NR}}\fr{\calE^2}{16M^3}
\right]\psi_{\vp=0},
\eeq
with the coefficient 
$c_{\textrm{NR}}=8M^3 \left( 4\pi\alpha_E- \frac{\mu^2}{4 M^3} \right)$, 
which should be compared with the energy shift in
\Eq{Eshift}. 
From integration over the Euclidean time
$\tau$, 
we can easily obtain the spin-averaged Green's function
\beq
G_{\textrm{NR}}(\tau)
&=&
\theta(\tau) \,
\textrm{exp}
\left[
-\fr{(Z\calE)^2\tau^3}{6M}+\fr{(Z\calE)^4\tau^5}{40M^3}
\right.
\nn \\
&&\left.
-\fr{(Z\calE)^6\tau^7}{112M^5}-c_{\textrm{NR}}\frac{\calE^2\tau}{16M^3}
\right].
\label{eq:nrqed_corr}
\eeq

Although we find the simple expression in 
\Eq{nrqed_corr} 
for the external field correlator, 
additional work is required to utilize this expression to fit lattice QCD correlation function data. 
Lattice correlators are relativistic and customarily traced over spinor space.
Given the form of a lattice correlation function
\beq
\mathbb{G}_{\alpha \beta}
(\vx, \tau)
=
\langle 0|
[\calO_\Psi]_\alpha (\vx, \tau)
[\calO_{\bPsi}]_\beta (\vec{0}, 0)
|0\rangle
,\eeq 
where
$\calO_\Psi$
is an interpolating operator for the spin-half hadron 
$\Psi$, 
we would trace over spin indices to arrive at the spin-averaged relativistic correlation function. 
In order to use the simple form of the correlator in the nonrelativistic limit, 
\Eq{nrqed_corr}, 
we need to perform the FW transformation which is both spin and coordinate dependent.  
In light of 
\Eq{positive_G},
the FW transformed lattice correlator is given by
\beq
\mathbb{G}_{\textrm{NR}}(\tau)
=
\sum_{\vx} 
\tr \left[
\calT_+(\vx,\tau)
\mathbb{G}
(\vx, \tau)
\gamma_4
\calT^\dagger_+(\vec{0},0)
\right]
\label{eq:transformation_mat} 
,\eeq
where the sum is carried out over the entire spatial lattice to project onto vanishing spatial momentum.
The matrix incorporating the FW transformation and positive energy projection is given by
\beq
\calT_+(\vx,\tau)
&=& 
\fr{1+\gamma_4}{2}
e^{M\tau} e^{iS_3(\vx,\tau)} e^{iS_2(\vx,\tau)} e^{iS_1(\vx,\tau)},
\eeq
with the generators
$S_i$ 
defined in Eqs.~(\ref{eq:S1})--(\ref{eq:S3}). 
In this particular example of a uniform electric field, 
the transformation takes the form
\beq
\calT_+(\vx,\tau)
&=&
\fr{1+\gamma_4}{2}
e^{ M \tau}
e^{ - \frac{i ( Z \calE \tau)^3}{6 M^3} \gamma_3}
e^{- \frac{ i \mu \calE}{4 M^2} \gamma_4 \gamma_3}
e^{ \frac{i Z \calE \tau}{2 M} \gamma_3}
\label{eq:T}
.\quad \, \,  \eeq
Thus to analyze lattice QCD data in a nonrelativistic QED framework, 
we require knowledge of the hadron mass 
$M$
and its magnetic moment 
$\mu$
in order to perform the FW transformation of the lattice correlator. 
Notice that a neutral particle is not immune to such difficulties;
its magnetic moment is required for the FW transformation.  
The hadron's mass can be determined from lattice data in vanishing electric field, 
whereas it is not clear how to determine the magnetic moment \emph{a priori}. 
Imagining the latter difficulty could be surmounted, 
one needs to write out correlator data for various spin components of the hadron's correlation function. 
The reason for this is that the determination of hadronic parameters is subject to uncertainty, 
and this uncertainty must also be propagated through the FW transformation, 
which ultimately involves weighting the spin components differently.  
These difficulties pose a challenge for a strict nonrelativistic QED analysis of lattice QCD correlation functions in external 
fields.
\footnote{
For the case of a charged scalar hadron
$\Phi$,  
there is an analogous complication in using a nonrelativistic QED framework to analyze lattice QCD correlation functions. 
The scalar correlation function calculated with lattice QCD is relativistic, 
and has the form
\beq
\mathbb{G}
(\vx, \tau)
= 
\langle 0|
\calO_{\Phi}(\vx,\tau)
\calO_{{\Phi}^\dagger}
(\vec{0},0)|0\rangle
\nn
,\eeq
where 
$\calO_\Phi$
is an interpolating operator for 
$\Phi$. 
To compute the nonrelativistic correlation function, 
one must account for the difference in normalization between the fields. 
Using the relation between the relativistic scalar field 
$\Phi$ 
and the nonrelativistic scalar field 
$\phi$, 
namely 
$\phi(x)=\calN(x)\Phi(x)$, 
we obtain
\beq
\mathbb{G}_{\textrm{NR}}(\vx, \tau)
=
\calN(\vx,\tau)
\mathbb{G} (\vx, \tau) 
\calN(\vec{0},0)
,\nn 
\eeq
with 
$\calN(\vx,\tau)=[4(M^2-\vD^2)]^{1/4} e^{M\tau}$.
In the particular case of a uniform electric field, 
the normalization factor alters the time dependence of the nonrelativistic correlation function, 
$\calN(\vx,\tau) \to [ 4 (M^2 + Z^2 \calE^2 \tau^2)]^{1/4} e^{M\tau}$. 
To compute this normalization factor, 
one needs to determine the mass of the scalar hadron, 
and appropriately propagate the uncertainty. 
This complication disappears for a neutral scalar hadron. 
}

\subsection{Relativistic correlation functions}
\label{sec:corr_rel}

The alternative to a nonrelativistic QED analysis of lattice data is an analysis employing a relativistic hadron theory. 
This is natural from the point of view of lattice QCD computations, 
as one avoids dealing with the FW transformation of data. 
The drawback of such an approach, 
however,  
is that the relativistic hadron theory does not have manifest power counting. 
Instead, 
one includes operators which become the most relevant in the nonrelativistic limit, 
and attempts to treat them in a fully relativistic fashion. 
Indeed this semi-relativistic philosophy is employed to write down the relativistic hadron theory in 
\Eq{relqed}.
The higher-derivative operators of the relativistic theory present an additional difficulty, 
as it becomes impossible to treat their contributions exactly. 
In~\Ap{reduced_theory}, 
we show how to handle the complication of higher-derivative operators for the special case of uniform external fields with vanishing hadron momentum.  
Fortunately the effects of these operators can be handled with these simplifying assumptions. 
As a result, 
our starting point will be the reduced hadronic theories given in 
Eqs.~(\ref{eq:RedE}) and (\ref{eq:RedB}). 
The former was previously adopted in a study of lattice QCD in uniform electric fields%
~\cite{Detmold:2010ts}. 
We first verify the correlators determined in that study, 
and provide a useful approximate form for the charged spinor correlator in a uniform electric field. 
The case of a uniform magnetic field is then taken up, 
where new results are obtained by generalizing Landau level projection to spin-half hadrons.

\subsubsection{Uniform electric field}

A charged particle in a uniform electric field does not have definite energy eigenstates. 
As a result, 
the correlation function will exhibit non-standard time dependence. 
For a charged spin-half hadron in a uniform electric field specified by  
the vector potential 
$A_\mu=-\calE \tau\delta_{\mu3}$, 
we have the reduced action density from \Eq{RedE}
\beq
\calL=
\bPsi_{\vec{p}=0}\left[
\gamma_4\partial_\tau+iZ\calE\gamma_3 \tau
+
\calM_\calE
-
\fr{\kappa\calE}{2M}
i\gamma_3\gamma_4
\right]
\Psi_{\vec{p}=0},
\nn \\
\label{eq:action_charged_E}
\eeq
where the fields are projected onto vanishing three-momentum $\vec{p}=0$, 
and 
$\calM_\calE$
includes the shift from the electric polarizability, 
$\calM_\calE = M + \frac{1}{2} 4 \pi \alpha_E \calE^2 + \cdots$, 
along with potentially higher-order effects from the electric field. 
Similar terms can be included by treating 
$\kappa$
as a function of 
$\calE^2$.

Following%
~\cite{Detmold:2010ts}, 
we consider boost-projected correlation functions
from which the anomalous magnetic moment and electric polarizability can be extracted using simultaneous fits. 
These boost-projected correlation functions have the form
\beq
G_\pm(\tau)=\int d \vx~ \textrm{tr}
\left[
K_\pm
\langle 0|\Psi(\vx,\tau)\bar{\Psi}(\vec{0},0)|0\rangle_\calE
\right].
\eeq
With our choice of gauge potential, 
the boost projection matrices are given by
\beq
K_\pm
=
1\pm i\gamma_3\gamma_4
,\eeq
up to their overall normalization. 
Note that this mixed momentum-time correlator does not put the hadron on shell 
because there are no energy eigenstates due to the 
explicit time dependence of the action density. 
Appealing to Schwinger's proper-time trick%
~\cite{Schwinger:1951nm,Tiburzi:2008ma}, 
we arrive at the integral representation for the correlation function%
~\cite{Detmold:2010ts}
\beq
G_\pm(\tau)
&=&
M
\left(1 \pm \fr{\kappa \calE}{2M^2}
\right)
\nn \\
&&
\times
\int_0^\infty ds
\sqrt{\fr{Z\calE}{2\pi\sinh Z\calE s}}
e^{-\fr{1}{2}\left[ \fr{Z\calE \tau^2}{\tanh Z\calE s}+E_{\pm}^2s
\right]},
\nn \\
\label{eq:rel_prop_uniform_E}
\eeq
where the parameters
$E_\pm$
depend on the boost projection, 
and are given by
\beq
E_{\pm}&=&\left[\calM_\calE^2-\fr{\kappa^2\calE^2}{4M^2}\pm Z\calE
\right]^{1/2}.
\eeq
These quantities are the analogue of a relativistic initial energy for the charged hadron, 
and there is a Zeeman-like splitting between boost-projected states arising from the Dirac magnetic moment, 
$\Delta E = E_+ - E_- \approx Z \calE / M$, 
for weak electric fields.

In the long-time limit, 
the correlator in 
\Eq{rel_prop_uniform_E}
does not follow a simple exponential decay. 
Because the integral representation of the correlation function results in cumbersome numerical  
fits to lattice QCD data, 
we consider the non-relativistic limit of the correlation function using nonrelativistic QED power counting.
The integral in 
\Eq{rel_prop_uniform_E} 
can be evaluated in the velocity expansion,
but 
\emph{without} 
performing the FW transformation. 
To this end,  
it is convenient to define the dimensionless parameters,
\beq
&\eta=E_{\pm} \tau \sim \calO(v^{-2}),&\nn \\
&\zeta=Z\calE \tau^2 \sim \calO(v^{-1}).
\eeq
Using the method devised in%
~\cite{Lee:2013lxa}, 
we find the  
$O(v^4)$
result for the external field correlation function
\beq
\label{eq:nonrel_prop_uniform_E}
G_\pm(\tau)
&=&
\left(1\pm\fr{\kappa\calE}{2M^2}\right)
e^{-\eta-\fr{\zeta^2}{6\eta}}
\Bigg[
1-\fr{\zeta^2}{4\eta^2}
\left(1-\fr{\zeta^2}{10\eta}\right)
\nn \\
&&
-\fr{\zeta^2}{4\eta^3}\left(1-\fr{5\zeta^2}{8\eta}
+\fr{17\zeta^4}{280\eta^2}-\fr{\zeta^6}{800\eta^3}\right)
\Bigg].
\eeq
This approximation for the correlation function sidesteps the need for numerical integration. 
While it has a somewhat involved form, 
it only depends on the hadronic parameters 
$M$, 
$\kappa$, 
and
$\alpha_E$, 
just as the un-approximated proper-time integral in 
\Eq{rel_prop_uniform_E}.

The results for a neutral spin-half hadron can be derived by performing the integral in 
\Eq{rel_prop_uniform_E} 
analytically in the limit of vanishing charge. 
The resulting relativistic correlation function of a boost-projected neutral spin-half hadron can be written simply as
\beq
G_\pm(\tau)
=
\left(
1\pm\fr{\kappa \calE}{2M^2}
\right)
e^{-E_\calE \tau},
\eeq
where 
$E_\calE=\sqrt{\calM_\calE^2 -\fr{\mu^2\calE^2}{4M^2}}$, 
with the relation
$\mu = \kappa$
for a neutral particle. 
This result also agrees with the approximate form of the integral derived in 
\Eq{nonrel_prop_uniform_E}, 
when evaluated at vanishing electric charge.

\subsubsection{Uniform magnetic field}

While the relativistic propagator for a charged point-like particle has long been known%
~\cite{Schwinger:1951nm}, 
the analogous result for a composite spin-half particle has thus far not been investigated. 
We derive this propagator starting from the reduced action in 
\Eq{RedB}, 
which properly accounts for the Dirac and anomalous magnetic moment terms.  

For a uniform magnetic field, 
we use the gauge choice
$A_\mu=-Bx_2\delta_{\mu1}$,
for which 
the 
$x_2$ component of momentum for a charged spin-half hadron is not a good quantum number. 
In fact, 
a correlation function which is averaged over all space will receive contributions from an infinite tower of Landau levels. 
After projection onto vanishing good components of momentum,
$p_1=p_3=0$, 
we arrive at the effective action density 
\beq
\calL
&=&
\bar{\psi}_{p_1=p_3=0}
\Bigg[
\gamma_4 \partial_\tau 
+ 
\gamma_2 \partial_2 
+
iZB\gamma_1 x_2
\nn \\
&& \phantom{spacings}
+
\calM_B
+
\fr{\kappa B}{2M}i\gamma_1\gamma_2
\Bigg]
\psi_{p_1=p_3=0},
\eeq
where 
$\calM_B=M- \fr{1}{2}4\pi\beta_M B^2 + \cdots$ 
includes the shift from the magnetic polarizability.
Note that the absence of explicit time dependence implies the existence of energy eigenstates. 
As a consequence, 
the extreme long-time behavior of the Euclidean correlation function will exhibit an exponential decay governed by the 
ground-state energy. 
Using the Schwinger proper-time trick to invert the action,
we find the propagator can be written in the form
\begin{widetext}
\beq
G_B(\tau,x_2)
&=&
\fr{1}{2}
\left[
-
\gamma_4\fr{\partial}{\partial\tau}
-
\gamma_2\partial_2
+
\calM_B
-
iZB\gamma_1 x_2
-
\fr{\kappa B}{2M}i\gamma_1 \gamma_2
\right]
\int_0^\infty 
\fr{ds}{\sqrt{2\pi s}}
e^{-\fr{\tau^2}{2s}-\fr{1}{2}s \calM_B^2}
\langle x_2 | e^{ - s \calH} | 0 \rangle
\nn \\
&&
\times 
\Bigg[
\left(
\calS_+ \calP_+ 
e^{\frac{Z B s}{2}}
+ 
\calS_- \calP_-
e^{-\frac{Z B s}{2}}
\right)
e^{\frac{\kappa B \tau}{2M}}
+
\left(
\calS_+ \calP_- 
e^{\frac{Z B s}{2}}
+ 
\calS_- \calP_+
e^{-\frac{Z B s}{2}}
\right)
e^{-\frac{\kappa B \tau}{2M}}
\Bigg]
,\label{eq:charged_corr_B}
\eeq
\end{widetext}
where 
$\calH$ 
is the auxiliary quantum mechanical Hamiltonian in
 $(x_2,\partial_2)$-space, 
see
\Eq{Ham}, 
and
$\langle x_2 | e^{ - s \calH} | 0 \rangle$
is the corresponding harmonic oscillator propagator, 
with 
$s$
playing the role of Euclidean proper time. 
We have simplified the spin structure by employing spin and parity projection matrices, 
which have the form
\beq
\mathcal{S}_{\pm}= \frac{1}{2} (1\mp i\gamma_1\gamma_2), 
\quad 
\text{and}
\quad 
\mathcal{P}_\pm= \frac{1}{2} (1\pm\gamma_4)
,\eeq
respectively. 
As we will see, 
the latter no longer project onto positive and negative energy states. 
In Minkowski space, 
the exponential factors in the last line of 
\Eq{charged_corr_B}
are phase oscillations that contain precession frequencies. 
With anomalous and Dirac magnetic moments, 
there are different precession frequencies: 
the former precesses in time, while the latter precesses in proper time. 
Notice that for odd parity,  
the two counterrotate.

The rather complicated form for this propagator arises from the implicit sum over contributions from the tower of Landau levels. 
The energy splitting between adjacent Landau levels is characterized by 
$\Delta E=|ZB|/M$ 
for weak external fields; 
and, 
in this limit, 
standard lattice spectroscopy will become considerably challenging.  
The correlation function in the long-time limit will still suffer from 
significant excited-state contamination,
now due to Landau levels rather than excited hadronic states.  
To improve the signal, 
we isolate the ground state by projecting the correlator
onto the lowest Landau level as suggested in%
~\cite{Tiburzi:2012ks}
\beq
[\calG^{n=0}_B
(\tau)]_{\alpha \beta}
=
\int d\vec{x}~\psi^{*}_{n=0}(x_2)
\langle 0|
\Psi_\alpha(\vec{x},\tau)
\bPsi_\beta
(\vec{0},0)|0\rangle_B,\nn\\ 
\eeq
where $\psi_0(x)$ is the ground-state harmonic oscillator wave-function,
which has the form
\beq
\psi_{0}(x)=e^{-\fr{1}{2}|ZB|x^2}
,\eeq 
up to normalization. 
In principle, 
we can project the correlator onto an arbitrary 
Landau level 
$n$
using the corresponding coordinate wave-function, 
$\psi_{n}(x)$.

It turns out that for a given Landau level, 
there are four distinct positive-energy eigenstates 
(and four negative-energy eigenstates). 
As can be anticipated from the form of the propagator in 
\Eq{charged_corr_B}, 
these eigenstates can be disentangled by using spin and parity projectors. 
Notice it would be incorrect to refer to the latter as energy projectors. 
The four different spin and parity projected correlators take the form
\beq
\calG^n_{(\pm_1,\pm_2)}(\tau)
&\equiv&
\tr
\left[
\mathcal{S}_{\pm_1} 
\mathcal{P}_{\pm_2} \,
\calG^n_B(\tau)
\right]
\nn \\
&=&
\mathcal{Z}^n_{(\pm_1,\pm_2)}
e^{-E^n_{(\pm_1,\pm_2)}\tau},
\label{eq:proj_charged_corr_B}
\eeq
and each has a falloff which is a simple exponential in Euclidean time. 
We use subscripts to denote the correlation of signs with the spin and parity projectors. 
The four amplitudes of the exponentials are given by
\beq
\mathcal{Z}^n_{(\pm_1,\pm_2)}
&=&
\fr{1}{2}\left[
\fr{\calM_B}{\sqrt{\calM_B^2
\mp_1 
ZB+|ZB|(2n+1)}}\pm_2 1
\right].
\nn \\
\label{eq:Zs}
\eeq
The energies of the four eigenstates are given by 
\beq
E^n_{(\pm_1,\pm_2)}
=
\sqrt{\calM_B^2 \mp_1 ZB+|ZB|(2n+1)} 
\mp_1 
\left(
\pm_2
\fr{\kappa B}{2M}
\right)
.\nn \\
\label{eq:Es}
\eeq

The number of energy eigenvalues for a charged particle with anomalous magnetic moment is somewhat surprising.  
The algebraic form of the energy eigenvalues for negative-parity states, 
moreover, 
shows that states exist in which the Dirac and anomalous magnetic moments are essentially antiparallel. 
Despite these surprising features, 
however, 
notice that the negative-parity amplitudes, 
$\mathcal{Z}^n_{(\pm, -)}$ 
in 
\Eq{Zs},
both scale as 
$Z B / M^2$
in the weak field limit. 
In other words, 
only the positive-parity correlators survive the nonrelativistic limit, 
and the Dirac and anomalous moments are consequently aligned. 
As a result, 
any experimental signature of such parity-odd states is considerably suppressed. 
In principal, 
nevertheless, 
the parity-odd correlation functions could be 
obtained from lattice QCD calculations using the background field method 
provided the magnetic field is not sufficiently small compared to the hadron mass 
$M$.

Let us focus on the parity-even correlation functions with energy eigenvalues given by 
$E^n_{(\pm,+)}$
in
\Eq{Es}. 
When the hadron's electric charge 
$Z$
vanishes, 
we obtain the correct correlation functions of a neutral spin-half hadron, 
\beq
G_{\uparrow \downarrow}(\tau)
=
e^{
-\left(
\calM_B\mp\fr{\mu B}{2M}
\right)
\tau}
,\eeq
where we have used the equality 
$\mu = \kappa$
for a neutral particle, 
and the arrows denote the whether the spin is aligned or anti-aligned with the magnetic field. 
On the other hand, 
when the charge is non-vanishing but the anomalous magnetic moment vanishes, 
$\kappa=0$, 
we recover the textbook relativistic Landau levels%
~\cite{Itzykson:1980rh}.
In this case,  
sectors of different parity become degenerate, 
and the amplitudes simply combine in the parity average.

When the field is sufficiently weak, 
$|ZB|\ll M^2$, 
we can expand the positive-parity energies in the form
\beq
E_{\uparrow \downarrow}
&=&
M
+ 
\frac{|ZB|}{M}
\left(
n
+
\frac{1}{2}
\right)
\mp
\frac{\mu B}{2M}
- 
\frac{1}{2} 4 \pi \beta_M B^2
\nn \\
&& 
-
\frac{Z^2 B^2}{8 M^3}
\left[ 1 + ( 2 n + 1)^2 \right]
\pm
\frac{Z B | Z B |}{2 M^3} \left( n + \frac{1}{2} \right)
,
\nn \\
\eeq 
up to corrections of 
$O(B^3)$. 
The first two magnetic field-dependent terms are the generalization of the relativistic Dirac-Landau result to the case of a non-vanishing anomalous magnetic moment. 
Obviously the Landau energy arises from the charge, 
while the Zeeman splitting arises from the magnetic moment,
$\mu = Z + \kappa$. 
Our result contains the first relativistic correction to the Zeeman splitting
\beq
E_\uparrow
- 
E_\downarrow
= 
-\frac{B}{M}
\left[
\mu 
- 
Z
\frac{|Z B|}{M^2}
\left( n + \frac{1}{2} \right)
\right]
\label{eq:Z}
,\eeq
which depends on the Landau level number. 
After averaging over spin polarization, 
the energy depends on the magnetic polarizability
\beq
\frac{
E_\uparrow
+ 
E_\downarrow
}{2}
&=&
M 
+ 
\frac{|ZB|}{M}
\left(
n
+
\frac{1}{2}
\right)
\left[
1 - \frac{|ZB|}{2 M^2} \left( n + \frac{1}{2} \right)
\right]
\nn \\
&&
- 
\frac{1}{2} 
\left( 
4 \pi \beta_M 
+ 
\frac{Z^2}{4 M^3}
\right)
B^2
\label{eq:shift}
,\eeq
however, 
the Landau energy produces the leading magnetic-field dependence. 
The magnetic polarizability is just one of three contributions, 
moreover, at second order in the magnetic field strength. 
The competing terms are the first relativistic correction to the nonrelativistic Landau energy, 
and the pole term.
\footnote{
We choose this terminology for how the shift arises diagrammatically in the relativistic theory. 
In that context, 
the last term of \Eq{shift} arises from a tree-level diagram with an intermediate-state hadron propagator. 
The hadron propagates between two Dirac magnetic moment interactions with the external field, 
$\sim Z B / 2 M$.
In the nonrelativistic limit, 
what remains of the intermediate-state propagator is a factor of
$1/M$. 
} 
The latter has a natural explanation in terms of nonrelativistic QED matching.  
If we separate off the effect from Landau levels, 
the nonrelativistic QED operator with coefficient
$c_{A_1}$
in 
\Eq{nrqed}
gives rise to the 
$B^2$
shift in the charged hadron's energy in a magnetic field.  
This shift is exactly that appearing in the last line of 
\Eq{shift} 
above.

\section{Summary}
\label{sec:summary}

Standard nonrelativistic QED for hadrons and the background field method are inconsistent. 
To resolve this inconsistency, 
we augment nonrelativistic QED with equation-of-motion operators. 
The augmented theory is shown in 
\Eq{nrqed}, 
and maintains a canonically normalized kinetic term. 
The introduction of such operators presents a problem 
because the low-energy coefficients of standard nonrelativistic QED 
are determined by computing amplitudes which describe physical processes. 
To resolve the inconsistency between nonrelativistic QED for hadrons and the background field method, 
we must determine the coefficients of equation-of-motion operators. 
We show that the underlying relativistic hadron theory 
in \Eq{relqed}
can be employed for this purpose. 
To derive nonrelativistic QED from the underlying relativistic theory, 
we iteratively employ the FW transformation.  
As a result, 
values for coefficients of the equation-of-motion operators are found in 
\Eq{eom}, 
and the inconsistency between nonrelativistic QED and the background field method is resolved.

As an effective field theory approach is entirely systematic and can often be quite simple, 
we investigate whether nonrelativistic QED can be employed in the analysis of lattice QCD correlation functions in background fields. 
Of course, 
we must augment nonrelativistic QED by equation-of-motion operators to achieve this. 
Unfortunately we run into difficulties with this approach. 
The correlation functions determined with lattice QCD are necessarily relativistic, 
and one must perform the FW transformation of lattice QCD data in order to analyze external field correlators using
a nonrelativistic QED framework. 
To perform this transformation, 
one requires knowledge of hadronic parameters,
and must accordingly propagate uncertainties through the transformation. 
As an example, 
we consider spin-half hadrons in a uniform electric field, 
and find they require the transformation matrix in 
\Eq{T}, 
which depends on the hadron's mass and magnetic moment. 
For a charged hadron, 
the transformation modifies the time dependence of the correlation function, 
which is undesirable. 
Consequently 
we argue that a hybrid framework is the most desirable for the analysis of lattice QCD data in external fields. 
The hybrid framework rests on an underlying relativistic hadronic theory, 
with operators chosen by their relevance in the nonrelativistic limit. 
In general, 
such theories are difficult to work with due to the lack of manifest power counting, 
however, 
for the simple cases of uniform electric and magnetic fields, 
we obtain the reduced theories in Eqs.~(\ref{eq:RedE}) and (\ref{eq:RedB}). 
The former theory was employed previously in a lattice QCD study in background electric fields%
~\cite{Detmold:2010ts}, 
and we find results completely consistent with that work. 
Using the velocity power counting of non-relativistic QED
(and without performing the FW transformation), 
we obtain an analytic form for boost-projected correlation functions, 
\Eq{nonrel_prop_uniform_E}, 
that can be utilized to avoid numerical integration in fits to lattice QCD data. 
We also determine the form of the propagator for a charged spin-half hadron in a uniform magnetic field. 
In contrast to the point-like particle case, 
we include an anomalous magnetic moment term. 
The resulting propagator has a considerably intricate Euclidean time dependence, 
which can be made tractable by isolating Landau levels. 
In the weak field limit, 
moreover, 
the Landau levels will become closely spaced and cause standard lattice spectroscopy to fail. 
The Landau-level-projected correlation function has a simple exponential falloff when 
additionally states of definite spin and parity are projected out. 
From expanding the energies in the weak magnetic field limit, 
we find the first relativistic correction to the Zeeman splitting, 
\Eq{Z}, 
and previously overlooked terms that contribute to the spin-averaged energy in a way identical to the magnetic polarizability, 
\Eq{shift}.

There are subsequent investigations that are natural in light of our analysis. 
One such investigation is the extension of the present work to one order higher in 
the nonrelativistic expansion. 
At 
$O(M^{-4})$, 
one encounters the so-called spin polarizabilities of hadrons%
~\cite{Ragusa:1993rm}. 
For a lattice QCD determination of these quantities%
~\cite{Engelhardt:2011qq,Lee:2011gz}, 
one must understand how correlation functions depend on spin polarizabilities in a relativistic hadron theory, 
and the degree to which equation-of-motion operators are important for the case of non-uniform electric and magnetic fields. 
Another line of investigation concerns implementing the Landau level projection advocated here. 
The lattice discretization will modify the coordinate wave-functions of Landau levels, 
and one must study how well a given level can be projected out of the correlation function. 
Additionally one must go beyond the point-like hadron picture and incorporate dynamical information into the projection technique to optimize overlap with the lowest Landau level of an extended hadron. 
Nevertheless, 
a hybrid approach incorporating ideas from nonrelativistic QED will be essential for the success of background field methods in lattice QCD.

\begin{acknowledgments}
We gratefully acknowledge support for this work from the U.S.~National Science Foundation,
under Grant No.~PHY$12$-$05778$.
The work of BCT is additionally supported by a joint 
City College of New York--RIKEN/Brookhaven Research Center fellowship,
and a grant from the Professional Staff Congress of the CUNY. 
\end{acknowledgments}

\appendix

\section{Reduced relativistic theory for uniform fields}
\label{sec:reduced_theory}

In this appendix, 
we show that for special cases of uniform electric and magnetic fields the 
$C_4$--$C_6$ 
terms in the relativistic hadron theory, 
\Eq{relqed}, 
can effectively be eliminated by using field redefinitions.
The special external fields considered correspond to those used in practice by lattice QCD simulations, 
namely uniform external fields that point along one spatial direction. 
A further simplifying assumption is the projection onto vanishing spatial momentum,
inasmuch as permitted.
This procedure establishes the reduced hadronic theory in uniform electric fields that was assumed in%
~\cite{Detmold:2010ts}, 
as well as the analogous reduced hadronic theory in uniform magnetic fields.

We carry out the reduction in Minkowski space-time; 
the conversion to Euclidean space-time can easily be performed. 
To begin, 
we consider the 
$C_6$
term, 
which, 
for the case of uniform external fields, 
can be effectively replaced by
\beq
D_\rho \bPsi D^\rho \Psi
F^{\mu \nu} F_{\mu \nu}
\longrightarrow
-
\bPsi
D^2 
\Psi 
F^2
.\eeq
We then use the identity
\begin{equation}
-
D^\mu D_\mu
=
\slashed{D}\slashed{D}
+
\frac{Z}{2} 
F_{\mu \nu} \sigma^{\mu\nu}
,\end{equation} 
along with the field redefinition 
\beq
\Psi
=
\left[
1-\fr{C_6 F^2}{2M^5}
\left( 
i \slashed{D} + M
\right)
\right]
\Psi'
\label{eq:frd1}
,\eeq 
in order to absorb the effect of the  
$C_6$ term 
into lower-dimensional operators.
Specifically when written in terms of the 
$\Psi'$
field,  
the coefficient
$C_3$
in \Eq{relqed}
is replaced by
$C'_3$
which is given by
\beq
C'_3=C_3+C_6
.\eeq
Note that there are also field-dependent shifts of the remaining low-energy constants that lead to terms in the action of 
at least
$O(M^{-5})$, 
which accordingly have not been taken into account in our calculations above. 
Furthermore,
the effect of the field redefinition \Eq{frd1} on Green's functions is innocuous from the point of view of lattice QCD:
the overall normalization changes, 
and there is an additional delta-function contribution that is only relevant when the source and sink are at the same space-time point.

To reduce the remaining two terms in \Eq{relqed}, 
we utilize uniform external fields aligned with one particular axis, 
which is taken to be the 
$x_3$-axis.
Not surprisingly, 
we must adopt differing field redefinitions in the presence of electric and magnetic fields. 
For a uniform electric field arising from the gauge potential 
$A_\mu=-E t \delta_{\mu 3}$, 
the electromagnetic stress-energy tensor has the form, 
$T^\mu_\nu=\fr{1}{2}\textrm{diag}(E^2,-E^2,-E^2,E^2)$. 
After projecting the field onto vanishing three momentum 
$\vp=0$, 
the 
$C_4$ 
and 
$C_5$ 
terms can be written as
\beq
\fr{C_4}{2M^4}\bar{\Psi}'i\slashed{D}\Psi'E^2+\fr{C_5}{2M^5}\bar{\Psi}' D_\mu D^\mu\Psi'E^2
,\eeq
where 
$D_\mu = (\partial_0, 0,0, - i Z A_3)$ 
for this particular case. 
With this form, 
these operators can be eliminated by employing the field redefinition
\beq
\Psi'
=
\left[
1
-
\fr{C_4 E^2}{4 M^4}
-
\fr{C_5 E^2}{4 M^5}
\left( i \slashed D + M \right)
\right]
\Psi''
\label{eq:frd2}
,\eeq 
which results in the simple modification 
$C'_3 \to C''_3$,
where this new coefficient is given by
\beq
C''_3=C_3+C_6-\fr{1}{4}(C_4+C_5).
\eeq
Similar to our treatment of the 
$C_6$ 
term, 
we disregard 
any field-dependent shifts of low-energy constants arising from terms in the action of  
$O(M^{-5})$
and higher.
The field redefinition in \Eq{frd2}, 
moreover, 
does not alter Green's functions in a way that is relevant to lattice QCD computations.

For the case of a uniform magnetic field, 
we choose the gauge potential 
$A_\mu=-Bx_2\delta_{\mu 1}$ 
in order to orient the field along the 
$x_3$-axis. 
Consequently
the electromagnetic stress-energy tensor takes the form 
$T^\mu_\nu=\frac{1}{2}\textrm{diag}(B^2,-B^2,-B^2,B^2)$. 
For this choice of gauge, 
we project the field onto vanishing good components of momentum, 
namely 
$p_1=p_3=0$. 
In this setup, 
the remaining two terms then have the form
\beq
\fr{C_4\bar{\Psi}'i(\gamma^0 D_0-\vga\cdot\vD)\Psi'B^2}{2M^4}
+\fr{C_5\bar{\Psi}'(D^0D_0-\vD^2) \Psi'B^2}{2M^5}
,\nn\\
\label{eq:BC4C5}
\eeq 
where now the gauge covariant derivative reduces to
$D_\mu = ( \partial_0, - i Z A_1, \partial_2, 0)$
for this case. 
In contrast to a uniform electric field, 
we cannot completely eliminate the
$C_4$
and
$C_5$ 
terms upon redefining the field. 
To proceed further, 
we note that the square of the kinematic momentum operator 
$\vD$ 
is related to the 
quantum mechanical harmonic oscillator
Hamiltonian in 
$(\partial_2,x_2)$-space, 
namely
\beq
\calH = - \frac{1}{2} \partial_2^2 + \frac{1}{2} (Z B x_2)^2 = - \frac{1}{2} \vD^2
\label{eq:Ham}
,\eeq  
where the corresponding eigenstates are Landau levels having the energies
$E_n = |ZB| \left(n+\frac{1}{2} \right)$. 
For the low-lying Landau levels, 
it is thus sensible to treat the kinematic momentum operator perturbatively in the weak field limit, 
$B\ll M^2$. 
For this reason, 
we need not worry about removing the terms in \Eq{BC4C5} involving the kinematic momentum upon field redefinition. 
With this observation in mind, 
we employ
\beq
\Psi'
=
\left[
1
-
\fr{C_4 B^2}{4M^4}
-
\fr{C_5 B^2}{4M^5} \left( i \gamma^0 D_0 + M \right)
\right]
\Psi'''  
\label{eq:frd3}
,\quad \eeq 
for the case of a uniform magnetic field. 
The result of this field redefinition is the modification of the coefficient
$C'_3 \to C'''_3$, 
with
\beq
C'''_3=C_3+C_6+\fr{1}{4}(C_4+C_5).
\eeq
By assuming the lowest Landau level can be isolated, 
the pieces of 
\Eq{BC4C5} 
remaining after field redefinition lead to higher-order effects. 
Such effects are simple for the 
$\vD^2$ 
part of the 
$C_5$
term because the Landau levels are eigenstates of this operator. 
For the 
$\vga\cdot\vD$
piece present in the 
$C_4$
term, 
however,
the perturbative corrections to the Green's function are more complicated, 
but fortunately are beyond the order we are working.

Finally we note that the field redefinition employed in 
\Eq{frd3}
will generally affect the Green's function through the action of the time derivative. 
In the spectral representation of the Green's function, 
each Landau level will be accompanied by an energy-dependent factor, 
however, 
this will not be problematic provided the lowest Landau level is isolated. 
In this case, 
the field redefinition amounts to changing the overall normalization of the contribution from the 
lowest Landau level.

To summarize: 
starting with the relativistic theory given in \Eq{relqed}, 
we obtain the reduced theory valid at 
$O(M^{-3})$ 
for a charged spin-half hadron 
in a uniform external electric field
\beq
\calL=\bPsi\left[i\slashed{D}-M+\fr{\kappa}{2M}\sigma_{\mu\nu}F^{\mu\nu}+\frac{1}{2} 4 \pi\alpha_E E^2
\right]\Psi, \qquad
\label{eq:RedE}
\eeq
and analogously for a charged spin-half hadron in a uniform external magnetic field
\beq
\calL=\bPsi\left[i\slashed{D}-M+\fr{\kappa}{2M}\sigma_{\mu\nu}F^{\mu\nu}+ \frac{1}{2} 4\pi\beta_M B^2
\right]\Psi 
,\qquad
\label{eq:RedB}
\eeq
where the electric and magnetic polarizabilities have been identified as
\beq
\frac{1}{2} 4 \pi \alpha_E = - \frac{2 C''_3}{ M^3}, 
\quad 
\text{and}
\quad
\frac{1}{2} 4 \pi \beta_M = 2 \frac{C'''_3}{M^3}
.\eeq
As a result of the reduction procedure, 
the polarizabilities have been written as linear combinations of coefficients of the fully relativistic theory, 
and agree with those determined from matching with two-photon amplitudes, 
see
\Eq{pols}. 
In the case of a uniform electric field, 
the reduced theory in
\Eq{RedE}
is precisely that previously employed in%
~\cite{Detmold:2010ts}
(after continuation to Euclidean space-time), 
where higher-order effects were effectively subsumed into field-strength dependent coefficients
$\kappa$
and
$\alpha_E$.

\section{Charged spinor in a magnetic field on a Euclidean torus}
\label{sec:Euclidean_torus}

In this appendix, we investigate the two-point correlation function 
for a charged spin-half hadron in a magnetic field on a spatial Euclidean torus. 
After spin and parity projection, 
the spin-half case is entirely analogous to the scalar case, 
which was presented in%
~\cite{Tiburzi:2012ks}.
As a result, 
we obtain a simple generalization of the Landau level projection for charged spin-half particles on a torus.

In deriving the two-point correlation function for a charged spin-half hadron, 
we consider a continuous torus of length 
$L$ 
in each of the three spatial directions, 
and keep the temporal extent infinite. 
The gauge potential 
$A^{FV}_\mu =  - B x_2 \delta_{\mu 1}$ 
is only periodic up to a gauge transformation
\beq
A^{FV}_1(x+\hat{x}_2L)=A^{FV}_1(x)+\partial_1\Lambda(x),
\eeq
with the other two spatial directions periodic. 
The gauge transformation function has the form 
$\Lambda(x)=-BLx_1$. 
Accordingly a periodic matter field after the boundary gauge transformation must satisfy the relation
\beq
\Psi^{FV}(x+\hat{x}_2L)=e^{-iZBLx_1}\Psi^{FV}(x)
,\eeq
which is a so-called magnetic periodic boundary condition, 
see, 
for example%
~\cite{AlHashimi:2008hr}.
The matter field remains periodic in the remaining spatial directions. 
Consistency of these boundary conditions requires that the magnetic flux through 
the plane perpendicular to the magnetic field is quantized according to  
$ZBL^2=2\pi N_\phi$, 
where 
$N_\phi$ 
the magnetic flux quantum of the torus%
~\cite{'tHooft:1979uj,Smit:1986fn,Damgaard:1988hh}. 
Due to the fractional electric charges of quarks, 
the flux quantum
$N_\phi$
for a hadron will be related to the integer flux quantum of the down quark
$n_\phi$
through the relation 
$N_\phi = - 3 n_\phi$.

The ground-state wave function on a torus
can be obtained from summing images of the infinite volume wave-function
\beq
\psi_0^{FV}(\vec{x}_\perp)=\sum_{\nu_2 = - \infty}^\infty
\psi_0(x_2+\nu_2 L)e^{2\pi iN _\phi \nu_2 x_1/L}
\label{eq:psiFV}
,\eeq
in order to satisfy magnetic periodicity. 
The finite volume version of the Landau-level-projected correlation function in 
\Eq{proj_charged_corr_B} 
is given by
\beq
\calG^{FV}_{0, \uparrow \downarrow}(\tau)
&=&
\int_{-\frac{L}{2}}^{\frac{L}{2}}d\vec{x}_\perp 
\psi^{*FV}_0(\vec{x}_\perp)
\tr
\left[
\calS_{\pm} \calP_+ G^{FV}_B(\tau,\vec{x}_\perp)
\right],
\nn \\
\label{eq:GFV}
\eeq
where the correlation function on a torus
$G^{FV}_B(\tau,\vec{x}_\perp)$ 
can be written as a sum of magnetic periodic images of the infinite volume correlation function,
\beq
G^{FV}_B(\tau,\vec{x}_\perp)&=&\sum_{n_1,\nu_2} 
e^{2\pi i n_1 x_1 /L}e^{2\pi i N_\phi \nu_2 x_1/L}
\nn \\
&&
\times
G_B
\left(\tau,x_2+\nu_2 L-\fr{n_1}{N_\phi}L;-\fr{n_1}{N_\phi}L\right).
\nn \\
\eeq
Notice we use 
$\nu_2$
for the winding number, 
and 
$n_1$
for the momentum mode number. 
We can perform the 
compact integration over $x_1$, 
which results in a Kronecker delta on the momentum mode number,
$\delta_{n_1,N_\phi(\nu_2-\nu'_2)}$, 
where $\nu'_2$ denotes the winding number that arises from the 
ground-state wave-function. 
After performing the summation over 
$n_1$, 
and reindexing 
$\nu_2 \to - \nu_2 + \nu'_2$, 
we then obtain
\beq
\calG^{FV}_{0, \uparrow \downarrow}(\tau)
&=&
\sum_{\nu_2,\nu'_2} 
\int_{-\frac{L}{2}}^{\frac{L}{2}} dx_2
\, \psi^{*}_0(x_2+\nu'_2 L)
\nn \\
&&
\times
\tr
\left[
\calS_{\pm} \calP_+ G_B(\tau,x_2+\nu'_2 L; \nu_2 L)
\right].
\eeq
In analogy with the Poission summation formula, 
the compact integral over 
$x_2$ 
along with the infinite sum over 
$\nu'_2$ 
can be rewritten as an integral over the entire real line. 
This is accomplished by changing variables from 
$x_2$
to 
$x_2 - \nu'_2 L$. 
After performing the infinite integral over 
$x_2$, 
we arrive at
\beq
\calG^{FV}_{0, \uparrow \downarrow}(\tau)
=
\mathcal{Z} \,
\calG^{n=0}_{(\pm, +)}(\tau),
\eeq
where the overall constant 
$\mathcal{Z}$ 
is just an infinite sum over 
$\nu_2$, 
namely
$\mathcal{Z}=2 \sum_{\nu_2}e^{-\pi N_\phi \nu_2^2}$. 
Up to normalization, 
this is merely the finite volume ground-state wave-function at the source location,
$\vec{x}_\perp = 0$. 
The correlator 
$\mathcal{G}^{n=0}_{(\pm,+)}$
appearing above is the infinite volume correlator projected onto the lowest Landau level, 
which appears in \Eq{proj_charged_corr_B}. 
The time dependence of the finite volume correlator is thus a simple exponential falloff with Euclidean time.

In practice, 
it is useful to imagine the torus as extending asymmetrically from 
$0$ to $L$
rather than symmetrically from 
$- \frac{L}{2}$ to $\frac{L}{2}$. 
Due to the exponential falloff of the ground-state oscillator wave-function in space, 
moreover, 
only the first few magnetic periodic images need to be accounted for. 
This issue was discussed in considerable detail in%
~\cite{Tiburzi:2012ks}.
A highly efficient approximation to the ground-state Landau level's wave-function on a torus extending from 
$0$ to $L$
in each spatial direction is thus
\beq
\psi_0^{FV}
(\vec{x}_\perp)
&\approx&
\psi_0 (x_2)
+ 
e^{ - i Z B L x_1}
\psi_0(x_2 - L)
\nn \\
&&
+
e^{ i Z B L x_1}
\psi_0 (x_2 + L)
\nn \\
&&
+
e^{ - 2 i Z B L x_1}
\psi_0(x_2 - 2 L)
.\eeq
One only need be mindful that the projected correlator in \Eq{GFV}
requires the complex conjugate of this wave-function.

\bibliography{spinor_v2}

\end{document}